\documentclass[%
reprint,
superscriptaddress,
nobibnotes,
amsmath,amssymb,
aps,
pra,
showkeys,
]{revtex4-1}

\usepackage{graphicx}
\usepackage{bm}
\usepackage{amsfonts}
\usepackage{wasysym}
\usepackage{siunitx}

\begin{document}

\title{Application of a Zero-latency Whitening Filter \\
	to Compact Binary Coalescence Gravitational-wave Searches}

\author{Leo Tsukada}
\email{tsukada@resceu.s.u-tokyo.ac.jp}
\affiliation{
	Research Center for the Early Universe (RESCEU), Graduate School of Science, \\The University of Tokyo, Tokyo 113-0033, Japan
}%
\affiliation{
	Department of Physics, Graduate School of Science, The University of Tokyo, Tokyo 113-0033, Japan
}%

\author{Kipp Cannon}%
\email{kipp@resceu.s.u-tokyo.ac.jp}
\affiliation{
	Research Center for the Early Universe (RESCEU), Graduate School of Science, \\The University of Tokyo, Tokyo 113-0033, Japan
}%

\author{Chad Hanna}
\affiliation{The Pennsylvania State University, University Park, Pennsylvania 16802, USA}

\author{Drew Keppel}
\affiliation{California Institute of Technology, Pasadena, California 91125, USA}

\author{Duncan Meacher}
\affiliation{The Pennsylvania State University, University Park, Pennsylvania 16802, USA}

\author{Cody Messick}
\affiliation{The Pennsylvania State University, University Park, Pennsylvania 16802, USA}

\date{\today}

\begin{abstract}
	Joint electromagnetic and gravitational-wave (GW) observation is a major goal of both the GW astronomy and electromagnetic astronomy communities for the coming decade.  One way to accomplish this goal is to direct follow-up of GW candidates.  Prompt electromagnetic emission may fade quickly, therefore it is desirable to have GW detection happen as quickly as possible. A leading source of latency in GW detection is the whitening of the data. We examine the performance of a zero-latency whitening filter in a detection pipeline for compact binary coalescence (CBC) GW signals. We find that the filter reproduces signal-to-noise ratio (SNR) sufficiently consistent with the results of the original high-latency and phase-preserving filter for both noise and artificial GW signals (called ``injections''). Additionally, we demonstrate that these two whitening filters show excellent agreement in $ \chi^2 $ value, a discriminator for GW signals.
\end{abstract}

\maketitle


\section{\label{sec:1}Introduction}
	The main target of ground-based gravitational-wave detectors are GW signals from CBC and this design concept has been put on firmer grounds than ever since the first detection of GWs from binary black holes, GW150914 \cite{Abbott2016a}. In addition to the detection of GWs alone,  association of GW signals with electromagnetic (EM) counterparts will provide scientific benefits for multiple fields of astronomy. One of the most promising EM counterpart candidates is a short gamma-ray burst (SGRB) \cite{Metzher-Berger}. 
	  
	SGRBs are intense and non-repeating flashes of $\gamma$-rays with a duration of $\lesssim$\SI{2}{\second}. Their origin and mechanism have been investigated and, although no conclusive evidence has been discovered, it is widely accepted that SGRBs are emitted via a beamed relativistic jet from CBC events containing at least one neutron star (NS) \cite{Eichler-1989, Berger-SGRB}. The association of a GW signal with its SGRB counterpart will establish the CBC progenitor model. Furthermore, it will also enable us to improve localization precision of the GW source, which leads to identification of the host galaxy and the progenitor's local environment \cite{Metzher-Berger}.
	
	Despite the first successful GW detections by the advanced Laser Interferometer Gravitational-wave Observatory (aLIGO \cite{2015}),  there are still a couple of challenges to tackle toward the realization of a follow-up search with EM waves. One is the poor localizability of GW sources. This challenge can be addressed by a worldwide detector network on Earth. The aLIGO consists of two detectors at geographically separate sites: one in Hanford, WA, with 4 km arms; one in Livingston, LA, with 4 km arms. Furthermore, several other detectors are being constructed or upgraded. Virgo \cite{TheVirgo:2014hva} in Italy, for instance, will soon participate in a joint observation run with aLIGO, and KAGRA \cite{PhysRevD.88.043007} in Japan is now in the process of apparatus installation.
	
	Another challenge, more relevant to this work, is the latency of a data-analysis pipeline, specifically the GstLAL pipeline for this work. With respect to latency, the pipeline consists of five components: data calibration, data distribution, whitening process (explained in Section. \ref{subsec:whitening_filter}), trigger production, and alert distribution \cite{0004-637X-748-2-136,PhysRevD.95.042001}. The total latency is a little less than \SI{70}{\second} \cite{PhysRevLett.116.241103}, the first three are the major bottlenecks taking $\sim$\SI{30}{\second} in total.
	
	Because the most sensitive EM telescopes are unable to survey the entire sky continuously, alerting them to the occurrence at a CBC in a timely manner increases the chance of a successful association.　Theoretical work on several GRB models propose a vast range of the time lag between a GW emission and the onset of a following SGRB, from $<$\SI{10} {\second} \cite{GRB-GW-timelag} to \SIrange{e3}{e4}{\second} \cite{two-wind-model}. This motivates achieving alert latencies below \SI{10}{\second}.

	This work will focus on the whitening process among the components mentioned above and describe how the whitening filter can be optimized in terms of latency. Thereafter, we investigate the influence of this optimization on the original detectability of GW signals. In particular, we implement the zero-latency algorithm with a Finite-impulse-response (FIR) filter and compare the values of SNR and $ \chi^2 $ between the original (frequency-domain) and zero-latency algorithms. It should be noted that the zero-latency algorithm is applicable to general CBC data-analysis pipelines.
	
	This paper is organized as follows. In Section \ref{sec:2}, we give an overview of statistical method to analyze CBC GW signals and the comparison between the frequency-domain and zero-latency whitening algorithms. In Section \ref{sec:3}, we present the performance tests of the zero-latency whitening filter, including the comparison to the original whitening filter. Lastly, we conclude in Section \ref{sec:4}.
\section{\label{sec:2}Method}
	\subsection{Matched filtering and $ \chi^2 $ test}
		One statistic used to estimate the detection significance of GW signals is the signal-to-noise ratio (SNR), $ \rho $, computed using a  matched filter \cite{Allen2005}. For the calibrated strain data $ s(t) $ and a template waveform $ h(t) $, the output of the matched filter is formulated as shown below.
		\begin{align}
			z = (s(t), h(t)) \equiv 4 \int_{0}^{\infty} \frac{\tilde{h}^*(f) \tilde{s}(f)}{S_n(f)} df \label{eq:matched_filter}
		\end{align}
		where $ S_n(f) $ is the one-sided power spectral density of the detector strain noise. 
		$ \rho $ is defined as the output of the matched filter in the case of a normalized GW signal.
		\begin{align}
			\rho &\equiv \frac{z}{\sigma} \label{eq:snr}\\
			\text{where}\  \sigma^2 &\equiv 4 \int_{0}^{\infty} \frac{|\tilde{h}(f)|^2}{S_n(f)} df
		\end{align}

		Additionally, strain data contain noise transients which do not obey a Gaussian distribution and may accidentally produce high $ \rho $. Such non-Gaussian transient (referred to as ``glitches'') are distinct from real GW signals in that they do not have the morphology of the template $ h(t) $. Making use of this distinction, we employ another statistic, $ \chi^2 $, defined below, in order to distinguish the transients \cite{PhysRevD.95.042001}. The time-dependent SNR of data is compared with that expected from the a real signal using the auto-correlation function of the template at its time of peak amplitude, $ R(t) $. A $ \chi^2 $ value is computed for each trigger using the time-dependent SNR $ \rho(t) $, the peak SNR $ \rho_p $ at the timestamp of $ t_p $, the noise-weighted auto-correlation function of a template $ R(t) $.
		\begin{align}
			\chi^2 \equiv &\frac{1}{\mu}\int_{t_p-\delta t}^{t_p+\delta t} |\rho(t) - \rho_p R(t)|^2 dt\\ 
			\text{where} \quad &\rho(t) \equiv \frac{4}{\sigma} \int_{0}^{\infty} \frac{\tilde{h}^*(f) \tilde{s}(f)}{S_n(f)} \text{e}^{2\pi if(t - t_p)}df\\
			&R(t) \equiv \frac{4}{\sigma^2} \int_{0}^{\infty} \frac{|\tilde{h}(f)|^2}{S_n(f)} \text{e}^{2\pi if(t - t_p)}df
		\end{align}
		The factor $ \mu $ is to normalize the $ \chi $ value for a well-fit signal. The time window $ \delta t $ is a tunable parameter.
		
		Both SNR and $ \chi^2 $ values are used to derive a likelihood ratio necessary for ranking triggers \cite{likelihood-ratio}.

	\subsection{\label{subsec:whitening_filter}Frequency-domain whitening filter}
\begin{figure}[t]
				\begin{center}
					\includegraphics[width=3in]{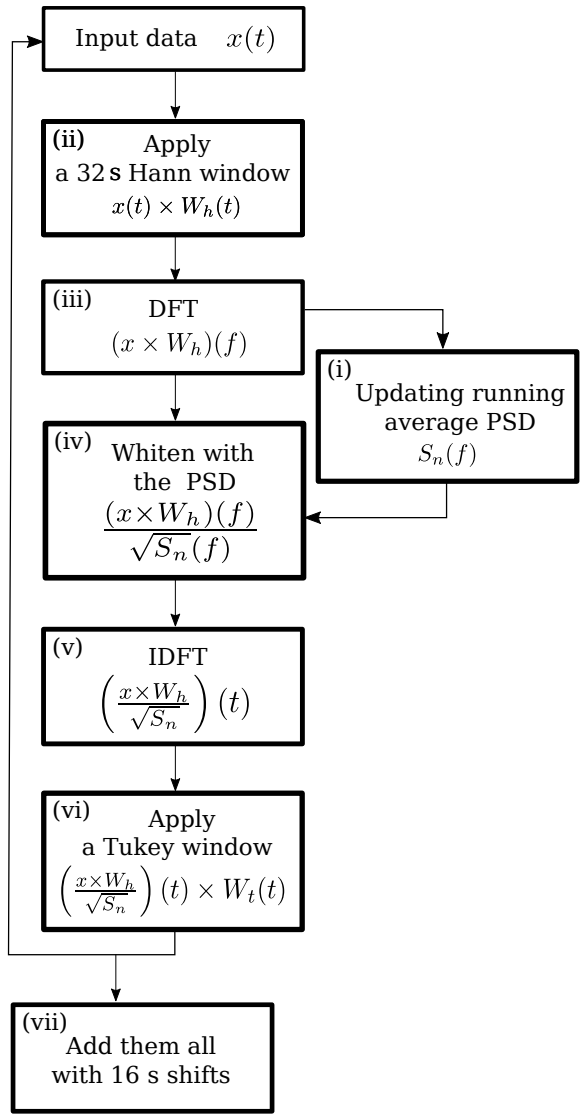}
				\end{center}
				\caption{\label{fig:current_algorithm}A schematic diagram of the frequency-domain whitening algorithm. Each numbered process corresponds to the numbers mentioned in Section \ref{subsec:whitening_filter}. $ W_h(t) $ and $ W_t(t) $ represent Hann and Tukey window functions, respectively.}
			\end{figure}

			As can be seen in Eq.(\ref{eq:matched_filter}), a matched filter is interpreted simply as an inner product between $ \tilde{s}(f) $ and $ \tilde{h}^*(f) $ with the weight of $ 1 / \sqrt{S_n(f)} $ for each. This weighting process is called ``whitening'', named after the fact that the transformation ideally returns only white noise.
			Referring to Fig.\ref{fig:current_algorithm}, the power spectrum density (PSD) of \SI{32}{\second} chunks of input data is measured for the subsequent whitening (i). Since the discrete Fourier transform (DFT) processes a chunk of a time series as if it were a periodic infinite series, we apply a Hann window function (ii) to suppress discontinuities on the boundaries of each period. These discontinuities will lead to click noise in the whitened data and may produce fake GW signals. Furthermore, in order to prevent any remaining discontinuity on the boundaries, some additional samples around the \SI{32}{\second} time series have been filled with zeros after applying the hann window. Then, the spectrum of the Hann-windowed block is computed (iii) and the block is whitened with the PSD (iv). After the inverse DFT (v), we further apply a Tukey window (vi). The purpose of this windowing is to suppress time-domain leakage which appears through the whitening and IDFT.
			The above procedure is repeatedly applied for every \SI{32}{\second} block with a \SI{50}{\percent} overlap and the PSD is updated every \SI{16}{\second}. In the end, all whitened data chunks, each of which is separately processed, are added with a consecutive \SI{16}{\second} shift in order to output a continuous time series of the whitened data (vii).
			The algorithm's main drawback is latency. Since a \SI{32}{\second} block is processed all at once every \SI{16}{\second}, the latency depends on the sample's location in the block, and can be anywhere from \SIrange{16}{32}{\second}.

	\subsection{\label{subsec:0LateFIR}Time-domain whitening filter}
		\subsubsection*{FIR-filter-based algorithm}
\begin{figure}[t]
			\begin{center}
				\includegraphics[width=\columnwidth]{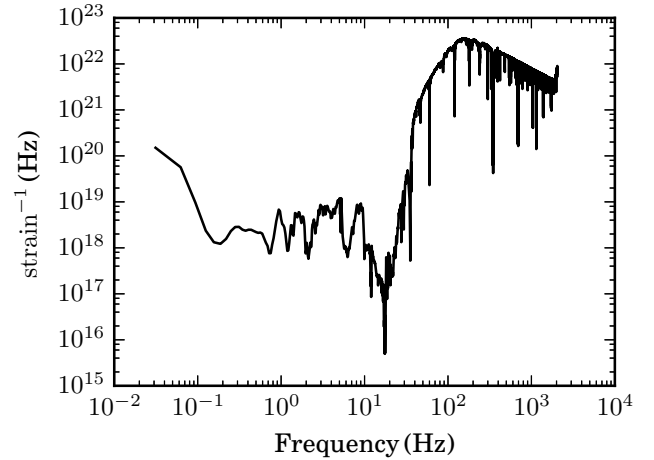}
			\end{center}
			\caption{\label{fig:amp_response} The amplitude response of the whitening filter. Due to the sample rate of \SI{4096}{\hertz}, the Nyquist frequency is \SI{2048}{\hertz}. The spectrum is computed simply from square-root of the inverse PSD.}
		\end{figure}

\begin{figure}[t]
			\begin{center}
				\includegraphics[width=\columnwidth]{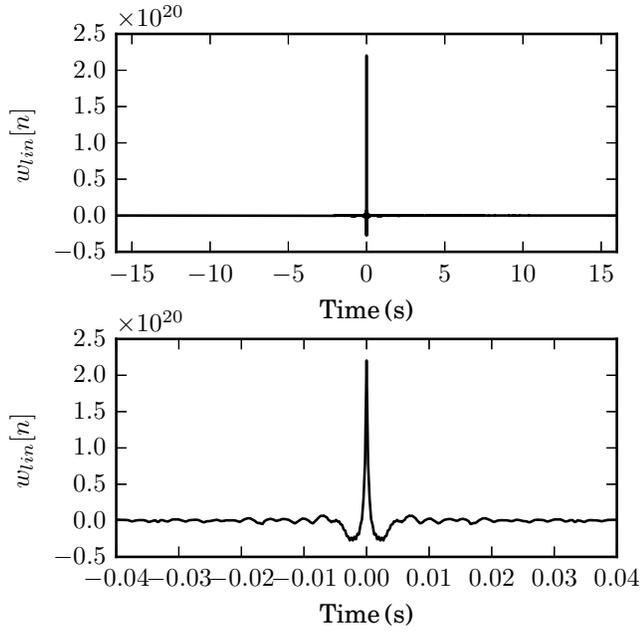}
			\end{center}
			\caption{\label{fig:kernel_linear_phase} The FIR of the original whitening filter. This is symmetric about its peak, a feature of a linear-phase filter. The peak in the middle leads to the latency of \SI{16}{\second}.}
		\end{figure}

			Here, we present an alternative FIR-filter-based algorithm to the frequency-domain whitening described above. Fig.\ref{fig:amp_response} shows the square root of the inverse PSD computed from the input LIGO strain data. This amplitude response has been employed to construct the FIR of a linear-phase filter shown in Fig.\ref{fig:kernel_linear_phase}. Therefore, it is possible to replace step (4) in Fig.\ref{fig:current_algorithm} with this FIR filter. It should be noted that due to its peak location, the FIR-filter-based algorithm still has a latency of \SI{16}{\second}.

		\subsubsection*{Zero-latency algorithm}
\begin{figure}[t]
			\begin{center}
				\includegraphics[width=\columnwidth]{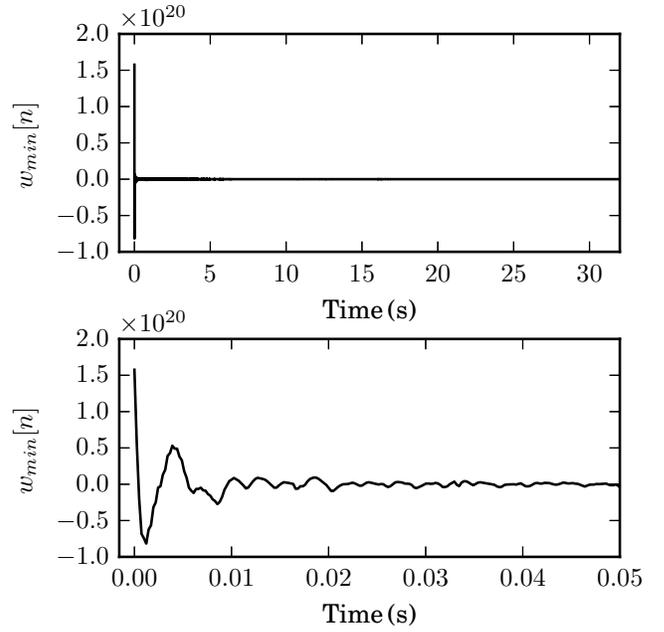}
			\end{center}
			\caption{\label{fig:kernel_min_phase} The impulse response of the zero-latency whitening filter. This FIR filter is purely causal.}
		\end{figure}

			According to the discussion in the previous paragraph, the peak of the filter must be moved to the left for the latency reduction. It is not possible to change the filter's latency without changing the whitening transformation. The result will be an approximation of the original filter. We adopt the approximation technique of N.Damera-Venkata \textit{et al}. \cite{840000} which approximately derives a phase response of a minimum-phase filter by applying the discrete Hilbert transform to the logarithm of a given magnitude response. Using more samples for the given magnitude response, one can more accurately approximate the magnitude response of the computed minimum-phase filter. The result is shown in Fig.\ref{fig:kernel_min_phase}. Fig.\ref{fig:kernel_min_phase} indicates that the FIR has its peak at the exact beginning, which is the reason why it is called a ``zero-latency whitening filter'' in this paper. As described in Section \ref{subsec:whitening_filter}, this whitening filter is equally applied to both of the templates and the strain data.

		\subsubsection*{Smooth PSD tracking}
			In order to replace the whole algorithm in Fig.\ref{fig:current_algorithm}, we have implemented an alternative method of the windowing process along with the whitening transformation. Specifically, we have allowed the PSD transition to occur continuously. Here, we have created a function which returns a linear combination of the newest and next newest filters during their transition as described by Eq.(\ref{eq:psdtracking}). The coefficient of the newer filter smoothly shifts from zero to one, sample by sample, according to a sinusoidal function. The zero-latency algorithm applies this function recursively any time a new whitening filter becomes available.
			\begin{align}
			\begin{split}
			s'(t) =
			\begin{cases}
			s_{old}(t) & (t < t_{up})\\
			&\\
			\cos^2\frac{2 \pi(t - t_{up})}{\varDelta t_{tr}}s_{old}(t)&\!\!\!\!\!+ \sin^2\frac{2 \pi(t - t_{up})}{\varDelta t_{tr}}s_{new}(t) \\
			\qquad & (t_{up} \leq t < t_{up} + \varDelta t_{tr})\\
			&\\
			s_{new} (t) & (t_{up}  + \varDelta t_{tr} \leq t)
			\end{cases}
			\end{split}
			\label{eq:psdtracking}
			\end{align}

		\begin{figure}[ht]
			\begin{center}
				\includegraphics[width=\columnwidth]{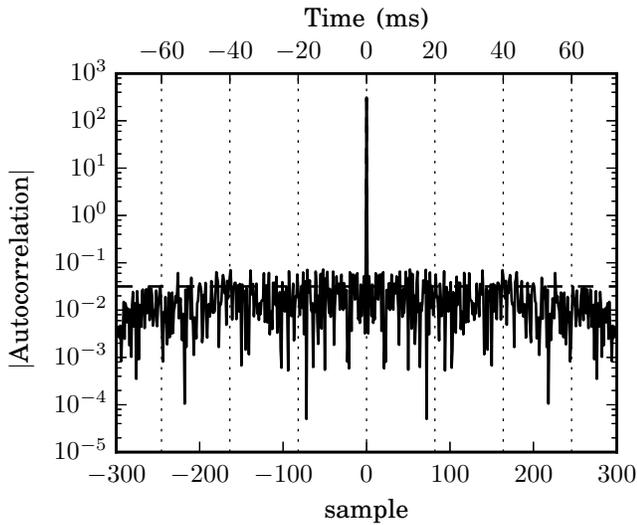}
			\end{center}
			\caption{\label{fig:autocorrelation_min_phase} Magnitude of auto-correlation of output stream from the zero-latency whitening filter. Technically, the absolute value is taken. It should be noted that this does not show any peak at the frequency of $\sim$ \SI{100}{\hertz}.}
		\end{figure}

		\begin{figure}[ht]
			\begin{center}
				\includegraphics[width=\columnwidth]{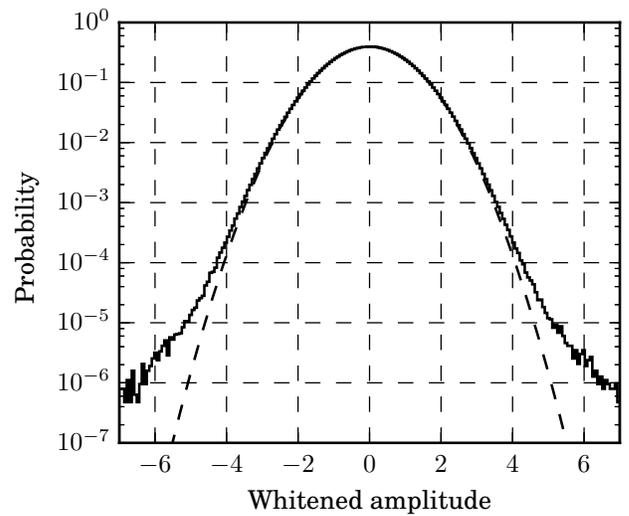}
			\end{center}
			\caption{\label{fig:histogram_min_phase} The logarithmic histogram of the output's amplitude.  The departure of the observed counts from the expected counts outside $(-5, 5)$ is due to the presence of non-Gaussian ``glitches'' in the interferometer data.}
		\end{figure}

		\clearpage

		\begin{figure}[ht]
			\begin{center}
				\includegraphics[width=\columnwidth]{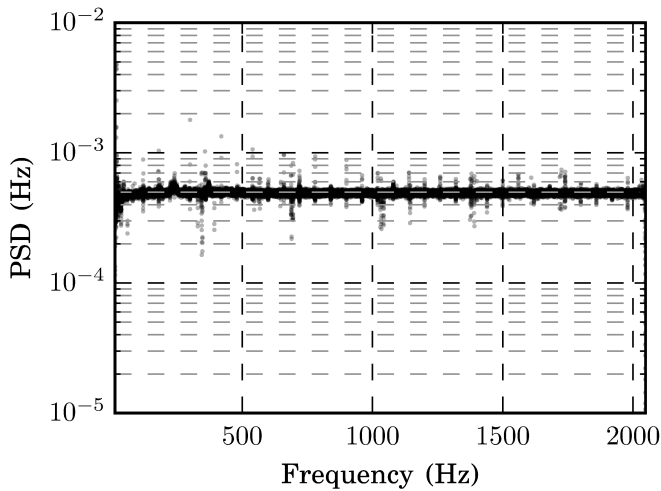}
			\end{center}
			\caption{\label{fig:white_psd} The averaged power spectral density of output stream from the zero-latency whitening filter. The data below \SI{12}{\hertz} are dropped off to ignore the effect of a high pass filter applied before the whitening filter.}
		\end{figure}

		where $ s'(t) $ is the resulting filter, $ s_{old, new} $ is the FIR of an older and newer filter respectively, $ t $ is the current time stamp, $ t_{up} $ is the time stamp when the PSD is updated and $ \varDelta t_{tr} $ is the duration of the filter transition. Particularly, we set $ \varDelta t_{tr} $ as \SI{0.125}{\second} in this work so that the transition timescale lies outside the frequency band of interest, which starts at \SI{10}{\hertz}. Note that this method is not unique in the application of a whitening filter but can be used for other time-dependent filtering.
\section{\label{sec:3}tests}
	Unlike the frequency-domain algorithm described in Section \ref{subsec:whitening_filter}, the zero-latency whitening filter does not conserve the phase of input data, which potentially harms the GW detectability. Therefore, it is necessary to demonstrate how significantly this change affects the resulting SNR and $ \chi^2 $.
	
	Here, we implement each of the zero-latency and frequency-domain whitening filters in the CBC gstLAL pipeline, which is compiled in LSC Algorithm Library (LAL) \cite{lal}. The pipeline scans LIGO strain data for any GW signal candidates (called ``triggers'') and computes SNR and $ \chi^2 $ for each trigger. In this work, we employ strain data from H1 with a duration of \SI{45,056}{\second} (from 08:25:23  to 20:56:19 UTC on 2005/11/27) during the fifth science run, called S5 \cite{0034-4885-72-7-076901}. A template bank is used spanning: component masses $ 3M_{\astrosun} \leq m_1, m_2 \leq 6M_{\astrosun} $; total mass $ 6M_{\astrosun} \leq M_{total} \leq 12M_{\astrosun} $; a minimal match of \SI{97}{\percent}; sampling frequency of \SI{2048}{\hertz}; non-spinning waveform to second post-Newtonian (PN) order. Along with statistical tests described below, we conduct two kinds of tests, namely, a noise-based and an injection-based test. In the noise-based test, the pipeline computes SNR and $ \chi^2 $ from the strain data with no GW signal. Therefore, all triggers in this test arise from detector noise which accidentally produce higher SNR than the threshold. On the other hand, the injection-based test requires artificial GW signals, and so we add injections to the same strain data. An injection is generated once every \SI{31.4}{\second} in the data, so the number of injections amounts to 1435. These two tests examine the agreement between the two whitening filters for noise and signals.

	\subsection{Statistical tests}
		Fig.\ref{fig:autocorrelation_min_phase} and Fig.\ref{fig:histogram_min_phase} show the auto-correlation and amplitude histogram created from an output stream of the zero-latency whitening filter. For comparison, expected curves are shown as dashed lines in the both figures, each of which indicates a delta-function with some variance and a Gaussian distribution respectively. Both plots show good agreement between the output and pure white noise. In particular, there is no apparent peak of the auto-correlation at around \SI{100}{\hertz}, at which the amplitude response shows its peak (See Fig.\ref{fig:amp_response}). Also, the averaged power spectral density is shown in Fig. \ref{fig:white_psd}. The spectrum is flatten throughout the shown frequency domain. Therefore, we conclude that the zero-latency algorithm sufficiently functions as a whitening filter.

	\subsection{\label{subsec:NoiseTest}Noise-based test}
\begin{figure*}[t]
			\begin{center}
				\includegraphics[clip, width=\textwidth]{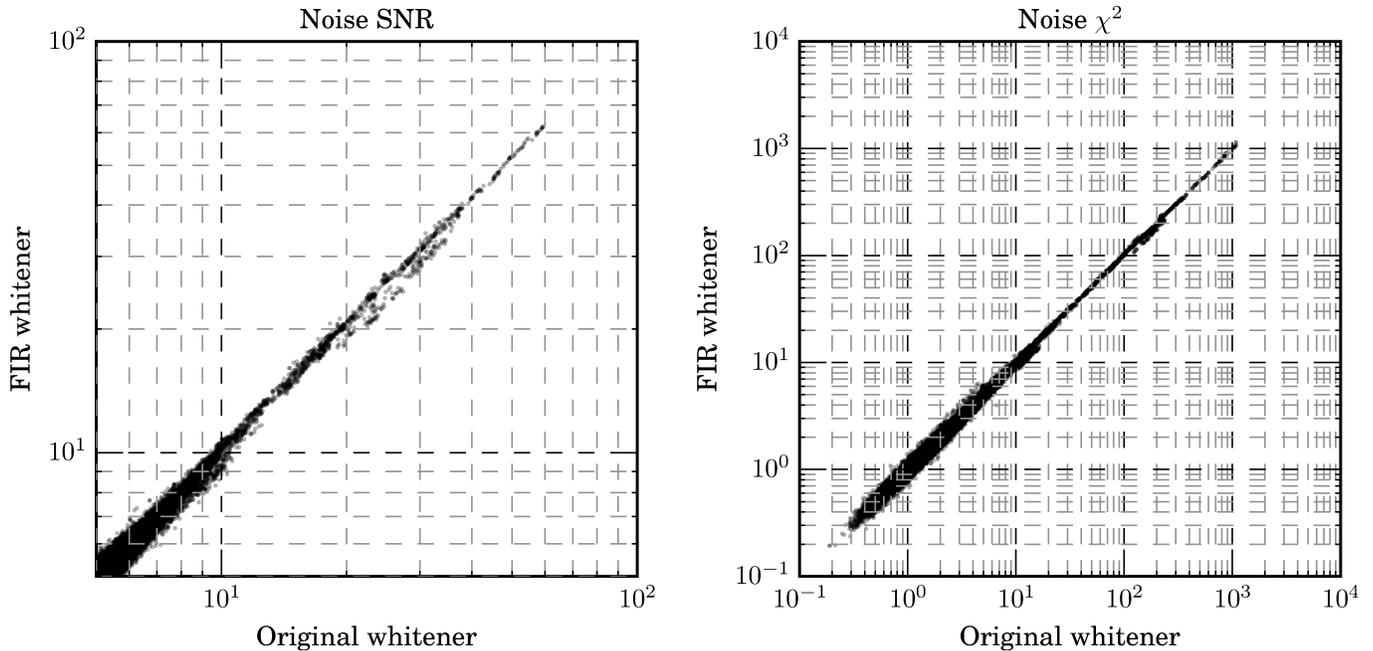}
			\end{center}
			\caption{\label{fig:snr_chisq_noise} Scatter plots of SNR and $ \chi^2 $ computed by the frequency-domain and zero-latency FIR whiteners for noise triggers. The value of SNR and $ \chi^2 $ range 5 $ - $ 60 and $ 0.2 - 10^{3} $ respectively. We have associated a pair of triggers within the end-time window of \SI{e-2}{\second} and with identical component masses, $ m_1 $ and $ m_2 $}
		\end{figure*}

		In Fig.\ref{fig:snr_chisq_noise}, we plot SNR and $ \chi^2 $ computed for each noise trigger by the zero-latency whitening filter versus the frequency-domain one. Here, we associate each counterpart by spotting a pair of triggers within the end-time window of \SI{e-2}{\second} and with identical component masses ($ m_1 $ and $ m_2 $). The two whiteners produced triggers with an SNR of 5 $ \sim $ 60 and a $ \chi^2 $ of $ 10^{-1} \sim 10^{3} $. Fig.\ref{fig:snr_chisq_noise} presents good agreement of both SNR and $ \chi^2 $ between the two whitening filters.

	\subsection{\label{subsec:InjectionTest}Injection-based test}
		\begin{figure*}[ht]
			\begin{center}
				\includegraphics[clip, width=\textwidth]{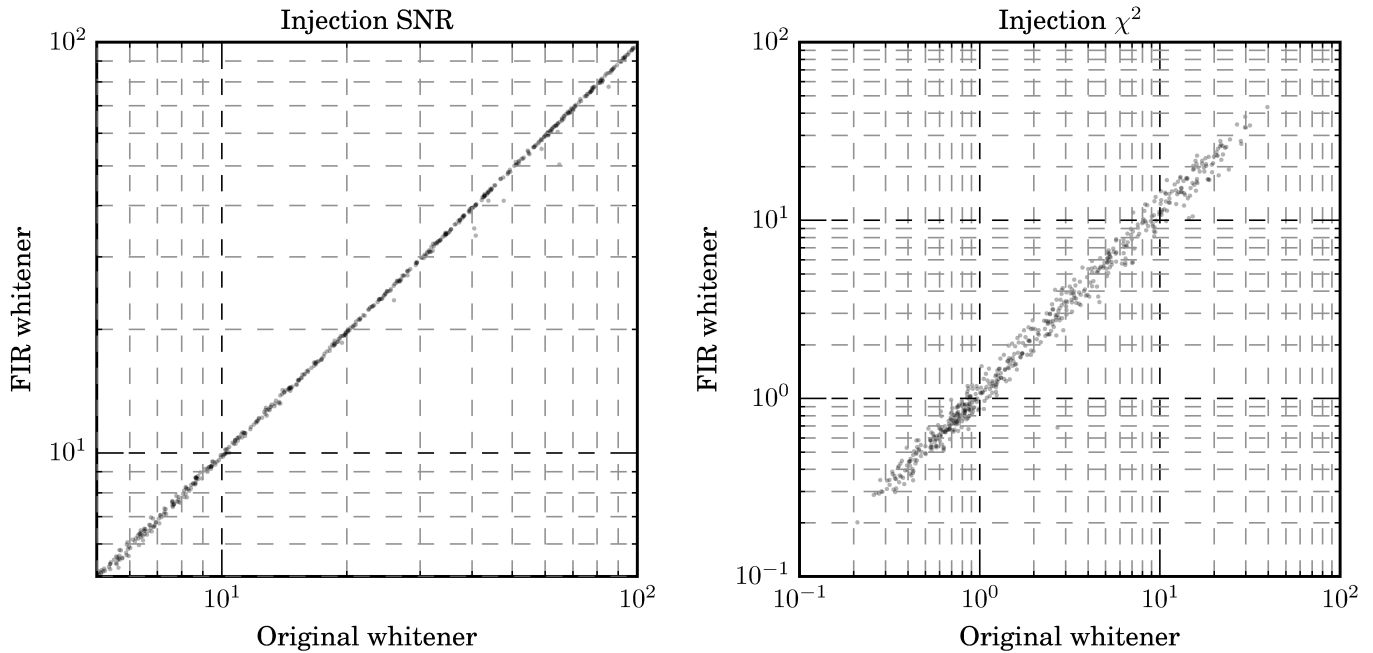}
			\end{center}
			\caption{\label{fig:snr_chisq_injection}Scatter plots of SNR and $ \chi^2 $ computed by the frequency-domain and zero-latency FIR whiteners for injection triggers. First, 1289 injections with the SNR of $ 5 - 10^4 $ were generated and truncated so that the maximum SNR would be 100. This is because triggers with higher SNR do not help to examine the critical discernibility of GW signals. Consequently, the values of SNR and $ \chi^2 $ range 5 $ - $ 100 and $ 0.3 - 400 $ respectively.}
		\end{figure*}
\begin{figure}[t]
			\begin{center}
				\includegraphics[clip, width=\columnwidth]{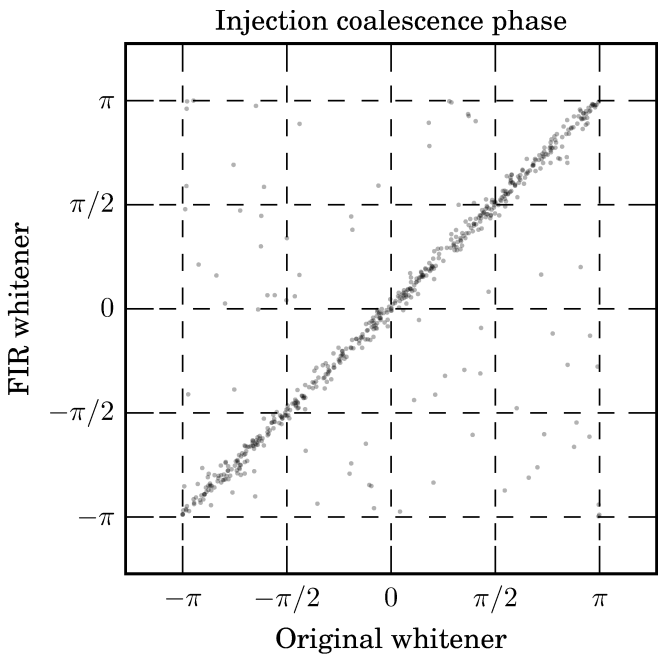}
			\end{center}
			\caption{\label{fig:phase_injection}The scatter plots of coalescence phase computed by the frequency-domain and zero-latency FIR whiteners for injection triggers. These triggers, whose SNR fall into the range from 5 to 100.}
		\end{figure}

		Fig.\ref{fig:snr_chisq_injection} shows SNR and $ \chi^2 $ computed with the zero-latency whitening filter versus the frequency-domain one in the presence of injections. Also, coalescence phase for every injection is shown in Fig.\ref{fig:phase_injection}. The coalescence phase is a phase of injection waveform at the coalescence time and determined by the ratio between cosine and sine components of a chosen template. In this test, we first simulate waveforms, based on a collection of parameters chosen randomly from a given probability distribution: $ m_1 $ and $ m_2 $ from a Gaussian distribution with the mean of $ 4.5M_{\astrosun} $ and the standard deviation of $ 0.5M_{\astrosun} $; cosine of the inclination chosen from a uniform distribution; non-spinning waveform to second post-Newtonian (PN) order with the cut-off frequency of \SI{30}{\hertz}. Next, the pipeline searches for and extracts injection by spotting the one \clearpage\hspace{-13pt} with the highest SNR among all located within \SI{1}{\second} of its true end time. After the trigger extraction, SNR, $ \chi^2 $ and coalescence phase of every injection trigger is recorded. In the end, only those with an SNR less than 100 are left to fit into the scatter plot, Fig.\ref{fig:snr_chisq_injection}. The above procedure is conducted for both whitening filters and we identify each counterpart by the end time of each injection. 

\begin{figure}[ht]
			\begin{center}
				\includegraphics[clip, width=\columnwidth]{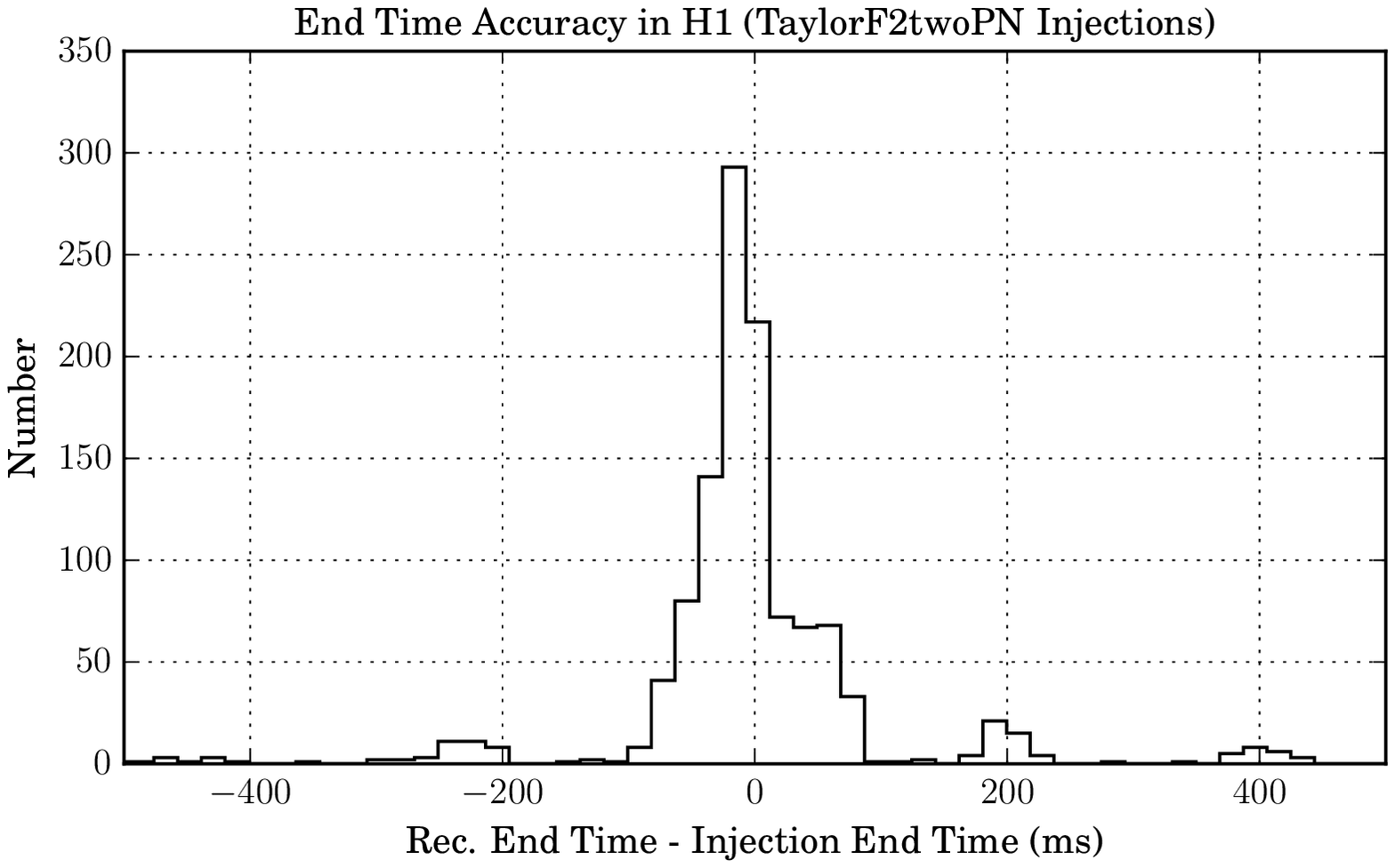}
			\end{center}
			\caption{\label{fig:injection_accuracy_fd} Histogram of the discrepancy between the true and estimated end time for each injection in case of the frequency-domain whitening filter. Each bin has the number of injection triggers whose end-time difference lies in the bin's range.}
		\end{figure}

\begin{figure}[ht]
			\begin{center}
				\includegraphics[clip, width=\columnwidth]{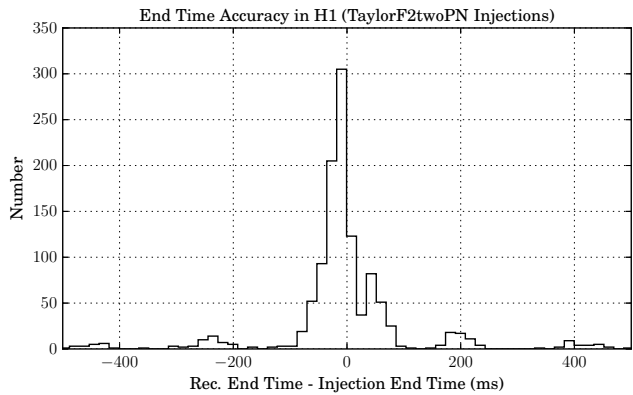}
			\end{center}
			\caption{\label{fig:injection_accuracy_td} Histogram of the discrepancy between the true and estimated end time for each injection in case of the zero-latency whitening filter. Each bin has the number of injection triggers whose end-time difference lies in the bin's range.}
		\end{figure}

		We have also conducted a consistency test for an injection's end time in the case of the two whitening filters. Fig.\ref{fig:injection_accuracy_fd}, \ref{fig:injection_accuracy_td} show histograms of the discrepancy between the true and estimated end time of each of the filters. In the both figures, the central peak has a tail width of $\sim$\SI{100}{\milli\second}, which is consistent with the typical tail width of the auto-correlation function of injected waveforms, suggesting that the pipeline properly generates and extract the injections from the triggers.
		
		As a result of the injection-based test, we find the SNR and $ \chi^2 $ computed with the zero-latency whitening filter to agree with those of the frequency-domain one. Although some triggers indicate that the new whitening filter slightly underestimates an SNR compared to the original one (See Fig.\ref{fig:snr_chisq_injection}), it will not harm the GW detectability since this case lies in the higher SNR regime.

\section{\label{sec:4}conclusion}
	We have applied and implemented an algorithm that optimizes latency for the whitening filter in the CBC data-analysis pipeline. Through statistical tests between the frequency-domain and zero-latency whitening filters, we have found that the two statistical values, SNR and $\chi^2$, are in sufficient agreement for both noise and injection triggers. As a result, we have achieved a \SI{16}{\second} latency reduction in the whitening process. It should be noted that this work has yielded the first confirmation that a zero-latency whitening filter can be employed in the data-analysis pipeline for CBC GW searches.

\begin{acknowledgments}
The authors are grateful to the LIGO lab for an offer of the S5 strain data. LIGO was constructed by the California Institute of Technology and Massachusetts Institute of Technology with funding from the National Science Foundation. This research was supported by the National Science Foundation through PHY-1454389 and ACI-1642391. Funding for this project was provided by the Charles E. Kaufman Foundation of The Pittsburgh Foundation. 
\end{acknowledgments}



\bibliography{ref}

\end{document}